\documentclass[twocolumn,aps,pra,superscriptaddress]{revtex4-2}
\usepackage{xcolor}
\usepackage{soul}

\usepackage{empheq}

\usepackage{mathtools,amsfonts,amsmath,amssymb,mathrsfs}
\usepackage{calrsfs}
\usepackage{exscale}
\usepackage{xfrac}
\usepackage{stmaryrd}
\usepackage{eucal}
\usepackage{graphicx} 
\usepackage{subfig}
\usepackage{graphics}
\usepackage{epsfig}
\usepackage{graphicx,color}           
\usepackage{color}
\usepackage{dcolumn}
\usepackage{bm}
\usepackage{float}
\usepackage{hyperref}
\usepackage[normalem]{ulem}  
\usepackage{multirow}  
\hypersetup{colorlinks,linkcolor={blue},citecolor={red},urlcolor={red}}
\graphicspath{{../graphics/}}

\begin{document}

\bibliographystyle{apsrev}
%
%
\title{ Theoretical realisation of a two qubit quantum controlled-not logic gate and a single qubit Hadamard logic gate in the anti-Jaynes-Cummings model}
%
%
\author{Christopher Mayero}
\affiliation{Department of Physics and Materials Science, Maseno University, Private Bag-40105, Maseno, Kenya}
%
%
\author{Joseph Akeyo Omolo}
\affiliation{Department of Physics and Materials Science, Maseno University, Private Bag-40105, Maseno, Kenya}     
\author{Onyango Stephen Okeyo}
\affiliation{Department of Physics and Materials Science, Maseno University, Private Bag-40105, Maseno, Kenya}
%
%
%
\begin{abstract}
We provide a theoretical scheme for realizing a Hadamard and a quantum controlled-NOT logic gates operations in the anti-Jaynes-Cummings interaction process. Standard Hadamard operation for a specified initial atomic state is achieved by setting a specific sum frequency and photon number in the anti-Jaynes-Cummings qubit state transition operation with the interaction component of the anti- Jaynes-Cummings Hamiltonian generating the state transitions. The quantum controlled-NOT logic gate is realised when a single atomic qubit defined in a two-dimensional Hilbert space is the control qubit and two non-degenerate and orthogonal polarised cavities defined in a two-dimensional Hilbert space make the target qubit. With precise choice of interaction time in the anti-Jaynes-Cummings qubit state transition operations defined in the anti-Jaynes-Cummings sub-space spanned by normalised but non-orthogonal basic qubit state vectors, we obtain ideal unit probabilities of success in the quantum controlled-NOT operations.  \\

\end{abstract}

\maketitle
%
%
\section{Introduction}
\label{sec:intro}

In quantum computers, quantum bits (qubits) \cite{nielsen2011quantum, scherer2019mathematics} are the elementary information carriers.  In such a computer, quantum gates \cite{nielsen2011quantum, scherer2019mathematics, deutsch1989quantum} can manipulate arbitrary multi-partite quantum states \cite{raussendorf2001one} including arbitrary superposition of the computational basis states, which are frequently also entangled. Thus the logic gates of quantum computation are considerably more varied than the logic gates of classical computation. In addition, a quantum computer can solve problems exponentially faster than any classical computer \cite{deutsch1992rapid} because by exploiting superposition principle and entanglement allows the computer to manipulate and store more bits of information than a classical computer.

In this paper we present a theoretical approach of realizing Hadamard and controlled-NOT (C-NOT) quantum logic gates which form a universal set for quantum computation \cite{barenco1995elementary, shi2002both, boykin2000new}.
The important discovery and proof of a conserved excitation number operator of the AJC Hamiltonian \cite{omolo2017conserved} now means that dynamics generated by the AJC Hamiltonian is exactly solvable, as demonstrated in the polariton and anti-polariton qubit (photospin qubit) models in \cite{omolo2017polariton, omolo2019photospins}. The reformulation developed in  \cite{omolo2017conserved, omolo2017polariton, omolo2019photospins}, drastically simplifies exact solutions of the AJC model which we shall apply in the present work.

We define the quantum C-NOT gate as that which affects the unitary operation on two qubits which in a chosen orthonormal basis in $\mathbb{C}^2$ gives the C-NOT operation obtained as

\begin{equation}
|a\rangle|b\rangle\rightarrow|a\rangle|a\oplus|b\rangle
\label{eq:theqcnot}
\end{equation}
where $|a\rangle$ is the control qubit, $|b\rangle$ is the target qubit and $\oplus$ indicates addition modulo 2 \cite{barenco1995conditional, scherer2019mathematics, nielsen2011quantum}. The C-NOT gate transforms superposition into entanglement thus acts as a measurement gate \cite{nielsen2011quantum, scherer2019mathematics,barenco1995conditional} fundamental in performing algorithms in quantum computers \cite{knill2002introduction}. Transformation to a separable state (product state) is realized by applying the C-NOT gate again. In this case, it is used to implement Bell measurement on the two qubits \cite{braunstein1992maximal}. 

We note here that the JC model has been applied extensively in implementing C-NOT and Hadamard gate operations. Domokos \textit{et al} (1995) \cite{domokos1995simple} showed that using induced transitions between dressed states, it is possible to implement a C-NOT gate in which a cavity containing at most one photon is the control qubit and the atom is the target qubit. Later, Vitali, D. \textit{et al} (2001)\cite{vitali2001quantum} proposed a scheme of implementing a C-NOT gate between two distinct but identical cavities, acting as control and target qubits respectively. By passing an atom prepared initially in ground state consecutively between the two cavities a C-NOT ($cavity\rightarrow{atom}$) and a C-NOT ($atom\rightarrow{cavity}$) is realised with the respective classical fields S. Saif, F. \textit{et al} (2001) \cite{saif2007engineering} presented a study of quantum computing by engineering non-local quantum universal gates based on interaction of a two-level atom with two modes of electromagnetic field in high Q superconducting cavity. The two-level atom acted as the control qubit and the two-mode electromagnetic field served as the the target qubit. In this letter, we apply an approach similar to that in \cite{saif2007engineering} where we implement a quantum C-NOT gate operation between two cavities defined in a two-dimensional Hilbert space spanned by the state vectors $|\mu_1\rangle=|1_A,0_B\rangle$ and $|\mu_2\rangle=|0_A,1_B\rangle$ as target qubits. Here $|\mu_1\rangle$ expresses the presence of one photon in mode A, when there is no photon in mode B, and $|\mu_2\rangle$ indicates that mode A is in vacuum state and one photon is present in mode B.  The control qubit in this respect is a two-level atom. The important difference with the approach used in \cite{saif2007engineering} is the model, i.e, while the initial absolute atom-field ground state $|g,0\rangle$  in the AJC interaction is affected by atom-cavity coupling, the  ground state $|g,0\rangle$ in the JC model \cite{saif2007engineering} is not affected by atom-cavity coupling. A similar result was determined independently in \cite{omolo2019photospins}. Further, with precise choice of interaction time in the AJC qubit state transition operations defined in the AJC qubit sub-space spanned by normalised but non-orthogonal basics qubit state vectors \cite{omolo2017polariton, omolo2019photospins}, C-NOT gate operations are realized between the two cavities.

The Hadamard gate also known as the Walsh-Hadamard gate is a single qubit gate \cite{scherer2019mathematics, nielsen2011quantum}. The Hadamard transformation is defined as 
 
\begin{equation}
\hat{H}=\frac{\hat{\sigma}_\textit{x}+\hat{\sigma}_\textit{z}}{\sqrt{2}}
\label{eq:mathhadamard}
\end{equation}
where it transforms atomic computational basis states $|e\rangle(|0\rangle)$, $|g\rangle(|1\rangle)$ into diagonal basis states according to
\begin{eqnarray}
\hat{H}|e\rangle\rightarrow\frac{|e\rangle+|g\rangle}{\sqrt{2}}\quad;\quad\hat{H}|g\rangle\rightarrow\frac{|e\rangle-|g\rangle}{\sqrt{2}}\nonumber\\
\hat{H}|0\rangle\rightarrow\frac{|0\rangle+|1\rangle}{\sqrt{2}}\quad;\quad\hat{H}|1\rangle\rightarrow\frac{|0\rangle-|1\rangle}{\sqrt{2}}
\label{eq:mathhadamard1}
\end{eqnarray}
 Vitali, D. \textit{et al} (2001)\cite{vitali2001quantum} showed that one qubit operations can be implemented on qubits represented by two internal atomic states because it amounts to applying suitable Rabi pulses. He demonstrated that the most practical solution on implementing one qubit operations on two Fock states is sending the atoms through the cavity. If the atom inside the cavity undergoes a $\frac{\pi}{2}$ pulse one realizes a Hadamard-phase gate. Saif, F. \textit{et al} (2001) \cite{saif2007engineering} also showed that it is possible to realise Hadamard operation by a controlled interaction between a two-mode high Q electromagnetic cavity field and a two-level atom. In his approach, the two-level atom is the control qubit, whereas the target qubit is made up of two modes of cavity field. Precision of the gate operations is realised by precise selection of interaction times of the two-level atom with the cavity mode. In this paper, we show that Hadamard operation in the AJC interaction is possible for a specified initial atomic state by setting a specific sum frequency and photon number in the anti-Jaynes-Cummings qubit state transition operation \cite{omolo2017polariton, omolo2019photospins}, noting that the interaction components of the anti-Jaynes-Cummings Hamiltonian generates state transitions.

The content of this paper is therefore summarised as follows. Section \ref{sec:model} presents an overview of the theoretical model. In sections \ref{sec:qcgate} and \ref{sec:hadgatelogic} respectively, implementation of a quantum C-NOT and Hadamard gates in the AJC interaction are presented. Finally section \ref{sec:conclusion} contains the conclusion. 
%
%
\section{The model}
\label{sec:model}
The quantum Rabi model of a quantized electromagnetic field mode interacting with a two-level atom is generated by the Hamiltonian \cite{omolo2017conserved} 

\begin{equation}
\hat{H}_R=\frac{1}{2}\hbar\omega(\hat{a}^{\dagger}\hat{a}+\hat{a}\hat{a}^{\dagger})+\hbar\omega_0\hat{s}_z + \hbar\lambda(\hat{a}+\hat{a}^{\dagger})(\hat{s}_++\hat{s}_-)
\label{eq:rabi1}
\end{equation}
noting that the free field mode Hamiltonian is expressed in normal and anti-normal order form $\frac{1}{2}\hbar\omega(\hat{a}^{\dagger}\hat{a}+\hat{a}\hat{a}^{\dagger})$.
Here, $\omega\hspace{1mm},\hspace{1mm}\hat{a}\hspace{1mm},\hspace{1mm}\hat{a}^{\dagger}$ are quantized field mode angular frequency, annihilation and creation operators, while $\omega_0\hspace{1mm},\hspace{1mm}\hat{s}_z\hspace{1mm},\hspace{1mm}\hat{s}_+\hspace{1mm},\hspace{1mm}\hat{s}_-$ are atomic state transition angular frequency and operators. The Rabi Hamiltonian in eq.~\eqref{eq:rabi1} is expressed in a symmetrized two-component form \cite{omolo2017conserved, omolo2017polariton, omolo2019photospins}

\begin{equation}
\hat{H}_R=\frac{1}{2}(\hat{H}+\hat{\overline{H}})
\label{eq:rabi2}
\end{equation}
where $\hat{H}$ is the standard JC Hamiltonian interpreted as a polariton qubit Hamiltonian expressed in the form \cite{omolo2017conserved}

\begin{eqnarray}
\hat{H}&=&\hbar\omega\hat{N}+2\hbar\lambda\hat{A}-\frac{1}{2}\hbar\omega\quad;\quad\hat{N}=\hat{a}^{\dagger}\hat{a}+\hat{s}_+\hat{s}_- \nonumber\\
\hat{A}&=&\alpha\hat{s}_z+\hat{a}\hat{s}_++\hat{a}^{\dagger}\hat{s}_-\quad;\quad\alpha=\frac{\omega_0-\omega}{2\lambda}
\label{eq:pham1}
\end{eqnarray}
while $\hat{\overline{H}}$ is the AJC Hamiltonian interpreted as an anti-polariton qubit Hamiltonian in the form \cite{omolo2017conserved}

\begin{eqnarray}
\hat{\overline{H}}&=&\hbar\omega\hat{\overline{N}}+2\hbar\lambda\hat{\overline{A}}-\frac{1}{2}\hbar\omega\quad;\quad\hat{\overline{N}}=\hat{a}\hat{a}^{\dagger}+\hat{s}_-\hat{s}_+\nonumber\\\hat{\overline{A}}&=&\overline{\alpha}\hat{s}_z+\hat{a}\hat{s}_-+\hat{a}^{\dagger}\hat{s}_+\quad;\quad\overline{\alpha}=\frac{\omega_0+\omega}{2\lambda}
\label{eq:antpham1}
\end{eqnarray}
In eqs.~\eqref{eq:pham1} and \eqref{eq:antpham1}, $\hat{N}$, $\hat{\overline{N}}$ and $\hat{A}$, $\hat{\overline{A}}$ are the respective polariton and anti-polariton qubit conserved excitation numbers and state transition operators.

Following the physical property established in  \cite{omolo2019photospins}, that for the field mode in an initial vacuum state only an atom entering the cavity in an initial excited state $|e\rangle$ couples to the rotating positive frequency field  component in the JC interaction mechanism, while only an atom entering the cavity in an initial ground state $|g\rangle$ couples to the anti-rotating negative frequency field  component in an AJC interaction mechanism, we generally take the atom to be in an initial excited state $|e\rangle$ in the JC model and in an initial ground state $|g\rangle$ in the AJC model. 

Considering the AJC dynamics, applying the state transition operator $\hat{\overline{A}}$ from eq.~\eqref{eq:antpham1} to the initial atom-field \textit{n}-photon ground state vector $|g,n\rangle$, the basic qubit state vectors $|\psi_{gn}\rangle$ and $|\overline{\phi}_{gn}\rangle$ are determined in the form (\textit{n}=0,1,2,....) \cite{omolo2019photospins}

\begin{equation}
|\psi_{gn}\rangle=|g,n\rangle\quad;\quad|\overline{\phi}_{gn}\rangle=-\overline{c}_{gn}|g,n\rangle+\overline{s}_{gn}|e,n+1\rangle
\label{eq:entsuptate}
\end{equation}
with dimensionless interaction parameters $\overline{c}_{gn}$, $\overline{s}_{gn}$ and Rabi frequency $\overline{R}_{gn}$ defined as

\begin{eqnarray}
\overline{c}_{gn}&=&\frac{\overline{\delta}}{2\overline{R}_{gn}}\quad;\quad\overline{s}_{gn}=\frac{2\lambda\sqrt{n+1}}{\overline{R}_{gn}}\quad;\quad\overline{R}_{gn}=2\lambda{\overline{A}_{gn}}\nonumber\\
\overline{A}_{gn}&=&\sqrt{(n+1)+\frac{\overline{\delta}^2}{16\lambda^2}}\quad;\quad\overline{\delta}=\omega_0+\omega
\label{eq:parameters}
\end{eqnarray}
where we have introduced sum frequency $\overline{\delta}=\omega_0+\omega$ to redefine $\overline{\alpha}$ in eq.~\eqref{eq:antpham1}.

The qubit state vectors in eq.~\eqref{eq:entsuptate} satisfy the qubit state transition algebraic operations

\begin{equation}
\hat{\overline{A}}|\psi_{gn}\rangle=\overline{A}_{gn}|\overline{\phi}_{gn}\rangle\quad;\quad\hat{\overline{A}}|\overline{\phi}_{gn}\rangle=\overline{A}_{gn}|\psi_{gn}\rangle
\label{eq:traans}
\end{equation}
In the AJC qubit subspace spanned by normalized but non-orthogonal basic qubit state vectors  $|\psi_{gn}\rangle$\hspace{1mm},\hspace{1mm} $|\overline{\phi}_{gn}\rangle$  the basic qubit state transition operator $\hat{\overline{\varepsilon}}_g$ and identity operator $\hat{\overline{I}}_g$ are introduced according to the definitions \cite{omolo2019photospins}

\begin{equation}
\hat{\overline{\varepsilon}}_g=\frac{\hat{\overline{A}}}{\overline{A}_{gn}}\quad;\quad\hat{\overline{I}}_g=\frac{\hat{\overline{A}}^2}{\overline{A}_{gn}^2}\quad\Rightarrow\quad\hat{\overline{I}}_g=\hat{\overline{\varepsilon}}_g^2
\label{eq:anttransop1}
\end{equation}
which on substituting into eq.~\eqref{eq:traans} generates the basic qubit state transition algebraic operations

\begin{eqnarray}
\hat{\overline{\varepsilon}}_g|\psi_{gn}\rangle&=&|\overline{\phi}_{gn}\rangle\quad;\quad\hat{\overline{\varepsilon}}_g|\overline{\phi}_{gn}\rangle=|\psi_{gn}\rangle\nonumber\\\hat{\overline{I}}_g|\psi_{gn}\rangle&=&|\psi_{gn}\rangle\quad;\quad\hat{\overline{I}}_g|\overline{\phi}_g\rangle=|\overline{\phi}_g\rangle
\label{eq:algop11}
\end{eqnarray}
The algebraic properties \hspace{0.5mm} $\hat{\overline{\varepsilon}}^{2k}=\hat{\overline{I}}_g$ \hspace{0.5mm} and \hspace{0.5mm} $\hat{\overline{\varepsilon}}^{2k+1}=\hat{\overline{\varepsilon}}_g$ \hspace{0.5mm}easily gives the final property  \cite{omolo2019photospins}

\begin{equation}
e^{-i\theta\hat{\overline{\varepsilon}}_g}=\cos(\theta)\hat{\overline{I}}_g-i\sin(\theta)\hat{\overline{\varepsilon}}_g
\label{eq:antialgprop}
\end{equation}
which is useful in evaluating time-evolution operators.

The AJC qubit Hamiltonian defined within the qubit subspace spanned by the basic qubit state vectors $|\psi_{gn}\rangle$ , $|\overline{\phi}_{gn}\rangle$ is then expressed in terms of the basic qubit states transition operators $\hat{\overline{\varepsilon}}_g$, $\hat{\overline{I}}_g$ in the form \cite{omolo2019photospins}

\begin{equation}
\hat{\overline{H}}_g=\hbar\omega(n+\frac{3}{2})\hat{\overline{I}}_g+\hbar\overline{R}_{gn}\hat{\overline{\varepsilon}}_g
\label{eq:antijch2}
\end{equation}
%
%
\section{Quantum c-not gate operations}
\label{sec:qcgate}

In order to realise a C-NOT quantum gate operation in this context, we take a two-level atom as the control qubit, which is defined in a two dimensional Hilbert space with $|e\rangle$ and $|g\rangle$  as basis vectors, where $|e\rangle$  expresses the excited state of the two-level atom and $|g\rangle$ indicates the ground state. Two non-degenerate and orthogonal polarized cavity modes $C_A$ and $C_B$ make the target qubit. The target qubit is defined in two-dimensional Hilbert space spanned by the state vector $|\mu_1\rangle=|1_A,0_B\rangle$, which expresses the presence of one photon in mode A, when there is no photon in mode B, and the state vector $|\mu_2\rangle=|0_A,1_B\rangle$, which indicates that mode A is in the vacuum state and one photon is present in mode B.

With reference to the AJC qubit state transition operator in eq.\eqref{eq:antialgprop}, lets first consider when an atom in ground state $|g\rangle$ enters an electromagnetic cavity with mode A in vacuum state and a single photon in mode B. The atom couples to the anti-rotating negative frequency component of the field mode undergoing an AJC qubit state transition. After the atom interacts with mode A for a time $t=\frac{\pi}{\overline{R}_{g0}}$, equal to half Rabi oscillation time, the driving field is modulated such that $\theta=\overline{R}_{g0}t=2\lambda\overline{A}_{g0}t=\pi$. Redefining \cite{omolo2019photospins}

\begin{equation}
\overline{\alpha}=\frac{\overline{\delta}}{2\lambda}=\frac{\omega_0-\omega+2\omega}{2\lambda}=\frac{\delta}{2\lambda}+\frac{\omega}{\lambda}=\alpha+\frac{\omega}{\lambda}
\end{equation}
and considering a resonance case where $\delta=\omega_0-\omega=0$ with $\lambda\gg\omega$,            $\overline{\alpha}$   becomes very small  thus $\theta=\lambda{t}=\frac{\pi}{2}$, since  $\overline{A}_{g0}=1$ determined from eq.~\eqref{eq:parameters}. The evolution of this interaction determined by applying the AJC qubit state transition operation in eq.~\eqref{eq:antialgprop} noting the definition of $\hat{\overline{I}}_g$ and $\hat{\overline{\varepsilon}}_g$  \citep{omolo2019photospins} in eq.~\eqref{eq:anttransop1} is of the form

\begin{equation}
e^{-i\theta\hat{\overline{\varepsilon}}_g}|g,0_A\rangle=\cos(\theta)|g,0_A\rangle-i\sin(\theta)|e,1_A\rangle
\label{eq:modeA}
\end{equation}
which reduces to

\begin{equation}
|g,0_A\rangle\rightarrow-i|e,1_A\rangle
\label{eq:flip1}
\end{equation}
We observe that the atom interacted with mode A and completed half of the Rabi oscillation, as a result, it contributed a photon to mode A and evolved to excited state $|e\rangle$. Now, after the interaction time, it enters mode B containing a single photon, interacting with the cavity mode as follows

\begin{equation}
-ie^{i\theta\hat{\overline{\varepsilon}}_e}|e,1_B\rangle=-i\cos(\theta)|e,1_B\rangle+\sin(\theta)|g,0_B\rangle
\label{eq:modeB}
\end{equation}
After an interaction with mode B for a time $t_1=2t$ such that $t_1=\frac{\pi(\overline{R}_{g0}+\overline{R}_{e1})}{\overline{R}_{g0}\overline{R}_{e1}}$, the driving field is modulated such that $\theta=\left(\frac{\overline{R}_{g0}\overline{R}_{e1}}{\overline{R}_{g0}+\overline{R}_{e1}}\right)t=\frac{\pi}{2}$ with $\overline{R}_{g0}=2\lambda{\overline{A}_{g0}}=2\lambda$ since $\overline{A}_{g0}=1$ and $\overline{R}_{e1}=2\lambda{\overline{A}_{e1}}=2\lambda$ since $\overline{A}_{e1}=1$. Therefore, $\lambda{t}=\frac{\pi}{2}$. The form of eq.~\eqref{eq:modeB} results into the evolution

\begin{equation}
-i|e,1_B\rangle\rightarrow|g,0_B\rangle
\label{eq:flip2}
\end{equation}
The results in  eq.~\eqref{eq:flip2} shows that the atom evolves to ground state and absorbs a photon initially in mode B. Therefore the atom clearly performs a swapping of the electromagnetic field between the two field modes by controlled interaction. 

When the atom in ground state $|g\rangle$, enters the electromagnetic cavity containing a single photon in mode A and mode B in vacuum state, the atom and the field interacts as follows

\begin{equation}
e^{-i\theta\hat{\overline{\varepsilon}}_g}|g,0_B\rangle=\cos(\theta)|g,0_B\rangle-\sin(\theta)|e,1_B\rangle
\label{eq:modeB2}
\end{equation} 
After an interaction with field mode B for a time $t=\frac{\pi}{\overline{R}_{g0}}$ equal to half Rabi oscillation time, the driving field is modulated such that $\theta=\overline{R}_{g0}t=\pi$, with $\overline{R}_{g0}=2\lambda{\overline{A}}_{g0}=2\lambda$ since $\overline{A}_{g0}=1$. Therefore $\theta=\lambda{t}=\frac{\pi}{2}$. The form of eq.~\eqref{eq:modeB2} results into the evolution

\begin{equation}
|g,0_B\rangle\rightarrow-|e,1_B\rangle
\label{eq:flip3}
\end{equation}
The atom then enters mode A containing one photon and interacts as follows

\begin{equation}
-e^{i\theta\hat{\overline{\varepsilon}}_e}|e,1_A\rangle=-\cos(\theta)|e,1_A\rangle-i\sin(\theta)|g,0_A\rangle
\label{eq:modeA2}
\end{equation}
After an interaction with the cavity mode for a time $t_1=2t$ such that $t_1=\frac{\pi(\overline{R}_{e1}+\overline{R}_{g0})}{\overline{R}_{e1}\overline{R}_{g0}}$ we obtain a driving field modulation $\theta=\left(\frac{\overline{R}_{e1}\overline{R}_{g0}}{\overline{R}_{e1}+\overline{R}_{g0}}\right)t=\frac{\pi}{2}$, with $\overline{R}_{e1}=2\lambda{\overline{A}_{e1}}=2\lambda$ since $\overline{A}_{e1}=1$ and $\overline{R}_{g0}=2\lambda{\overline{A}_{g0}}=2\lambda$ since $\overline{A}_{g0}=1$. Therefore $\theta=\lambda{t}=\frac{\pi}{2}$. The form of eq.~\eqref{eq:modeA2} results into the evolution

\begin{equation}
|e,1_A\rangle\rightarrow{i}|g,0_A\rangle
\label{eq:flip4}
\end{equation}
This shows that the atom evolves to ground state and performs a field swapping by absorbing a photon in mode A.

When the atom in excited state $|e\rangle$ enters mode A in vacuum state,that is target qubit $|\mu_2\rangle$, the atom propagates as a free wave without coupling to the field mode in vacuum state $|0\rangle$ \cite{omolo2019photospins}, leaving the cavity without altering the state of the cavity-field mode. A similar observation is made when the atom enters cavity B in vacuum state for the case of target qubit  $|\mu_1\rangle$.

From the results obtained, we conclude that the target qubit made up of the electromagnetic field remains unchanged if the control qubit, that is, the two-level atom, is initially in the excited state $|e\rangle$, while when the atom is in ground state $|g\rangle$, the cavity states $|0\rangle$ and $|1\rangle$ flip. We shall refer to this gate as the AJC C-NOT $(atom\rightarrow{cavity})$

\subsection{Probability of success of the C-NOT gate} 
\label{sec:cnotgate}
Success probability for the C-NOT gate is given by

\begin{equation}
P_s=1-(\sin^2(\theta_A)+\cos^2(\theta_A)\sin^2(\theta_B))
\label{eq:success1}
\end{equation}
In terms of the Rabi frequency we write eq.~\eqref{eq:success1} as

\begin{equation}
P_s=1-(\sin^2(\overline{R}_A\Delta{t_A})+\cos^2(\overline{R}_A\Delta{t_A})\sin^2(\overline{R}_B\Delta{t_B}))
\label{eq:success2}
\end{equation}
For the case in which the atom is in the ground state $|g\rangle$ and enters an electromagnetic cavity with mode A in vacuum state and a single photon of the field mode B

\begin{equation*}
\overline{R}_A=\overline{R}_{{g}{0}}=2\lambda{\overline{A}_{{g}{0}}}=2\lambda
\end{equation*}

\begin{equation*}
\Delta{t_A}=\frac{\pi}{\overline{R}_A}=\frac{\pi}{\overline{R}_{{g}{0}}}=\frac{\pi}{2\lambda}
\end{equation*}

\begin{equation*}
\overline{R}_B=\overline{R}_{{e}{1}}=2\lambda{\overline{A}_{{e}{1}}}=2\lambda
\end{equation*}

\begin{equation}
\Delta{t_B}=\frac{\pi}{2}\frac{(\overline{R}_A+\overline{R}_B)}{\overline{R}_A\overline{R}_B}=\frac{\pi}{2}\frac{(\overline{R}_{{g}{0}}+\overline{R}_{{e}{1}})}{\overline{R}_{{g}{0}}\overline{R}_{{e}{1}}}=\frac{\pi}{2\lambda}
\label{eq:first}
\end{equation}
substituting eq.~\eqref{eq:first} into eq.~\eqref{eq:success2} we obtain

\begin{equation}
P_s=1-(\sin^2(\pi)+\cos^2(\pi)\sin^2(\pi))=1
\label{eq:success3}
\end{equation}
a unit probability of success.

When the atom in ground state $|g\rangle$ enters an electromagnetic cavity containing a single photon in mode A, and mode B in the vacuum state

\begin{equation*}
\overline{R}_A=\overline{R}_{{e}{1}}=2\lambda{\overline{A}_{{e}{1}}}=2\lambda
\end{equation*}

\begin{equation*}
\Delta{t_A}=\frac{\pi}{2}\frac{(\overline{R}_A+\overline{R}_B)}{\overline{R}_A\overline{R}_B}=\frac{\pi}{2}\frac{(\overline{R}_{{e}{1}}+\overline{R}_{{g}{0}})}{\overline{R}_{{e}{1}}\overline{R}_{{g}{0}}}=\frac{\pi}{2\lambda}
\end{equation*}

\begin{equation*}
\overline{R}_B=\overline{R}_{{g}{0}}=2\lambda{\overline{A}_{{g}{0}}}=2\lambda
\end{equation*}

\begin{equation}
\Delta{t_B}=\frac{\pi}{\overline{R}_B}=\frac{\pi}{\overline{R}_{{g}{0}}}=\frac{\pi}{2\lambda}
\label{eq:second}
\end{equation}
substituting eq.~\eqref{eq:second} into eq.~\eqref{eq:success2} we obtain

\begin{equation}
P_s=1-(\sin^2(\pi)+\cos^2(\pi)\sin^2(\pi))=1
\label{eq:success4}
\end{equation}
a unit probability of success.

We observe that success probabilities depend mainly upon the precise selection of the interaction times of the two level atom with the successive cavity modes.

\section{Hadamard logic gate}
\label{sec:hadgatelogic}

To realise Hadamard operation in the AJC interaction, we apply the qubit state transition operation in eq.~\eqref{eq:anttransop1} and the general form in \cite{omolo2017polariton, omolo2019photospins}. In this respect, we define the Hadamard operation at sum frequency $\overline{\delta}=4\lambda$ and $n=0$ specified for an initial atomic state $|g\rangle$ as 

\begin{equation}
\hat{\overline{\varepsilon}}_g=\frac{1}{\overline{A}_{g0}}\left(2\hat{s}_z+\hat{a}\hat{s}_-+\hat{a}^\dagger\hat{s}_+\right)\hspace{5mm};\hspace{5mm}\overline{A}_{g0}=\sqrt{2}
\end{equation}
The initial atomic state $|g\rangle$ is rotated to

\begin{equation}
|g\rangle\rightarrow\frac{1}{\sqrt{2}}(|e\rangle-|g\rangle)
\label{eq:hadop1}
\end{equation}
Similarly, the Hadamard operation at sum frequency $\overline{\delta}=4\lambda$ and $n=1$ specified for an initial atomic state $|e\rangle$  is defined as \citep{omolo2019photospins}
\begin{equation}
\hat{\overline{\varepsilon}}_e=\frac{1}{\overline{A}_{e1}}\left(2\hat{s}_z+\hat{a}\hat{s}_-+\hat{a}^\dagger\hat{s}_+\right)\hspace{5mm};\hspace{5mm}\overline{A}_{e1}=\sqrt{2}
\end{equation}
The initial atomic state $|e\rangle$ is rotated to

\begin{equation}
|e\rangle\rightarrow\frac{1}{\sqrt{2}}(|e\rangle+|g\rangle)
\label{eq:hadop2}
\end{equation}
The Hadamard transformations in Eqs.~\eqref{eq:hadop1} and \eqref{eq:hadop2} realised in the AJC interaction process (AJC model) agree precisely with the standard definition in Eq.~\eqref{eq:mathhadamard1}.
%
\section{ Conclusion}
\label{sec:conclusion}
In this paper we have shown how to implement quantum C-NOT and Hadamard gates in the anti-Jaynes-Cummings interaction mechanism. We obtained ideal unit probabilities of success due to precise selection of interaction times during the C-NOT gate operations. We also realised efficient Hadamard operations through application of respective AJC qubit state transitions. 
%
%
\section*{Acknowledgment}

We thank Maseno University, Department of Physics and Materials Science for providing a conducive environment to do this work.

%
%
 \bibliographystyle{apsrev}

\bibliography{cmayero}

 \end{document}